\documentclass[%
twocolumn,
groupedaddress,
longbibliography,
 amsmath,amssymb,
 aps,
prb,
]{revtex4-2}
\usepackage{xcolor}
\usepackage{enumerate}
\usepackage{physics}
\usepackage[all]{xy}
\usepackage[T1]{fontenc}
\usepackage{amsmath,amsfonts,amssymb, braket}
\usepackage{wasysym}
\usepackage{comment}
\usepackage{enumerate}
\usepackage{graphicx}
\usepackage{mathtools}
\usepackage{ifthen}
\usepackage{tensor}
\usepackage{braket}
\usepackage{lipsum}
\usepackage{float}
\usepackage{standalone}
\usepackage{color}
\definecolor{myred}{RGB}{232,102,102}
\definecolor{myblue}{RGB}{187,187,255}
\definecolor{mygreen}{RGB}{34,139,34}
\definecolor{myorange}{RGB}{255,165,0}
\definecolor{OliveGreen}{RGB}{85,107,47}
\definecolor{NavyBlue}{RGB}{0,0,128}
\definecolor{mypink1}{rgb}{0.858, 0.188, 0.478}
\definecolor{mypink2}{rgb}{1, 0.188, 0.478}
\definecolor{mypink3}{rgb}{0.6, 0.188, 0.478}
\definecolor{mypink4}{rgb}{0.858, 0.195, 0.478}
\definecolor{mypink5}{rgb}{0.858, 0.3, 0.478}
\definecolor{mypink6}{rgb}{0.858, 0, 0.478}
\definecolor{mypink7}{rgb}{0.858, 0.188, 0.478}
\definecolor{mypink8}{rgb}{0.858, 0.188, 0.6}
\definecolor{mypink9}{rgb}{0.858, 0.188, 0.3}
\usepackage{tensor}
\newcommand{\be}{\begin{equation}}

\newcommand{\ee}{\end{equation}}
\newcommand{\ba}{\begin{aligned}}
\newcommand{\ea}{\end{aligned}}
\newcommand{\ignore}[1]{}
\usepackage[colorlinks,bookmarks=false,citecolor=blue,linkcolor=red,urlcolor=blue]{hyperref}

\usepackage{graphicx}% Include figure files
\usepackage{bm,bbm,bbold}% bold math

\usepackage{mathtools}

\begin{document}

\title{ Slow thermalization and subdiffusion in $U(1)$ conserving Floquet random circuits}

\author{Cheryne Jonay\footnotemark[1]}
\affiliation{Department of Physics, Stanford University, Stanford, CA 94305}

\author{Joaquin F. Rodriguez-Nieva\footnotemark[1]}
\thanks{Equal contribution}
\affiliation{Department of Physics, Stanford University, Stanford, CA 94305}

\author{Vedika Khemani}
\affiliation{Department of Physics, Stanford University, Stanford, CA 94305}

\date{\today}%

\begin{abstract}
Random quantum circuits are paradigmatic models of minimally structured and analytically tractable chaotic  dynamics. We study a family of Floquet unitary circuits with Haar random $U(1)$ charge conserving dynamics; the minimal such model has nearest-neighbor gates acting on spin 1/2 qubits, and a single layer of even/odd gates repeated periodically in time. We find that this minimal model is not robustly thermalizing at numerically accessible system sizes, and displays slow subdiffusive dynamics for long times.  We map out the thermalization dynamics in a broader parameter space of charge conserving circuits, and understand the origin of the slow dynamics in terms of proximate localized and integrable regimes in parameter space. In contrast, we find that small extensions to the minimal model are sufficient to achieve robust thermalization; these include (i) increasing the interaction range to three-site gates (ii) increasing the local Hilbert space dimension by appending an additional unconstrained qubit to the conserved charge on each site, or (iii) using a larger Floquet period comprised of two independent layers of gates. Our results should inform future studies of charge conserving circuits which are relevant for a wide range of topical theoretical questions. 
\end{abstract}

\maketitle

\section{Introduction \label{sec:intro}}

The dynamics of thermalization in isolated quantum systems is a topic of fundamental interest \cite{1991PRL_Deutsch, 1994PRE_Srednicki,2008Nature_Rigol,2016AnnPhys_ETHreview, 2016Science_Greiner}. Foundational questions on this topic have taken special relevance  with the advent of quantum simulators that can coherently evolve quantum states for long times~\cite{altman2021quantum, gross2017quantum, browaeys2020many, blatt2012quantum, kjaergaard2020superconducting}. Some key questions include identifying universal features in the dynamics of thermalizing systems, such as in the growth of entanglement entropy~\cite{2013PRL_Huseentanglementgrowth,2015JSM_Calabrese,2017PRB_wenwei, 2017PRX_quantumcircuits,     2019PRL_Rakovszky}, the propagation of quantum information~\cite{1972CMP_LiebRobinson,2006PRL_verstraete}, or the emergence of hydrodynamic transport~\cite{Khemani_2018, Rakovszky_2018, doyon2020lecture}. Likewise, understanding mechanisms by which systems evade thermalization or thermalize slowly, say due to localization~\cite{2015ARCMP_mbl,2019RMP_mbl,2019CRP_mbl} or quantum scarring~\cite{Serbyn-Papic2021_review,Moudgalya-Regnault2021_review,Chandran-Moessner2022_review} or  prethermalization~\cite{integrabilitybreaking3,2015PRX-babadi,2022PNAS_rodrigueznieva,2020NP_prethermalization,2020PRX_prethermalization, Mori-Saito2016_rigorous, Abanin-Huveneers2017_rigorous, RubioAbadal-Bloch2020}, is also an active area of research.  

Analytic approaches to studying the out-of-equilibrium dynamics of interacting quantum systems remain scarce. In this context, random quantum circuits have emerged as a powerful class of analytically tractable models that can capture universal features of quantum dynamics~\cite{Fisher_Vijay_CircuitsReview, 2017PRX_quantumcircuits, Nahum_2018,  von_Keyserlingk_2018}. Long studied in quantum information as models of quantum computation, quantum circuits are also the natural evolutions implemented by digital quantum simulators that are the subject of much experimental investigation~\cite{2018review_preskill, altman2021quantum, kjaergaard2020superconducting, 2019nature_googleqsupremacy}. 
A canonical example of a minimally structured random circuit retains only the ingredients of unitarity and locality, such that the time evolution of a set of a qudits on some lattice is generated by unitary gates acting on $r$ contiguous qudits, and drawn independently from the Haar measure at each position and time-step. 
Such minimally structured circuit models have no continuous symmetries or conservation laws (the discrete time-evolution means even energy is not conserved), and exhibit thermalization to an infinite temperature state, with universal ballistic spreading of quantum information en route to thermalization \cite{2017PRX_quantumcircuits, Nahum_2018,  von_Keyserlingk_2018}.

Random circuits can also be endowed with additional structure to achieve a range of interesting dynamical behaviors, with the simplest such extension being the addition of a single $U(1)$ conservation law of total spin~\cite{Khemani_2018, Rakovszky_2018}. In the minimal implementation of this conservation law, each local unitary gate assumes a block diagonal structure, with each symmetry block within each gate drawn independently from the Haar measure. We will refer to such gates as ``Haar random U(1) conserving" gates. 
With this change, the ballistic information spreading is (provably) accompanied by diffusive spin dynamics, and the diffusion constant can be analytically derived~\cite{Khemani_2018, Rakovszky_2018}. The dynamics can be made subdiffusive if the $U(1)$ conservation law is accompanied by other dipole or multipole symmetries, or if the system is endowed with additional kinetic constraints~\cite{2019PRX_qclocalization, Khemani_shattering_2020, Sala_fragmentation_2020,  Gromov_2020_fractonhydro, Morningstar_freezing_2020, 2021PRL_qcsubdiffusion}.

Another important extension of random circuit dynamics is obtained by considering Floquet dynamics with time-periodicity, so that layers of random unitary gates are repeated in time after a period of $T$ time-steps. In generic Floquet systems, the appropriate equilibrium state is the infinite temperature state~\cite{Lazarides_2014,D_Alessio_2014}. However, the periodicity in time and randomness in space opens up the possibility of many-body localization (MBL), which is absent in circuits that are random in time. Indeed, several families of structured random Floquet circuits have been shown to exhibit a many-body localized regime~\cite{ZhangKhemaniHuse_Floquet, S_nderhauf_2018, Morningstar-Huse2021_avalanches,Dibyendu1,Dibyendu2,Moudgalya_2021}. In this context, ''structured'' pertains to Floquet circuits generated by specific gate sets parameterized by tunable interaction and disorder strengths and, importantly, with a preferred local basis for the disorder such that the models look increasingly diagonal in this basis as the disorder strength is increased. Increasing disorder or decreasing interactions leads to MBL.

However, localization is generally not expected in \emph{minimally structured} Haar random Floquet models (without any conservation laws), in which neighboring spins are strongly coupled with Haar random gates, and there is no preferred local basis in which the disorder is strong. In line with this intuition, Ref.~\cite{Chan_2018_2} proved that local Haar random Floquet circuits are chaotic in the limit of infinite local Hilbert space dimension $q$ for each qudit (specifically, they proved that the spectral form factor derived from the Floquet unitary exhibits random matrix theory (RMT) behavior), while Ref.~\cite{S_nderhauf_2018} numerically showed thermalization even for spin 1/2 qubits. The derivation of chaos in the spectral form factor was also extended to random Floquet circuits with $U(1)$ charge conservation symmetry~\cite{Friedman_2019}, in which case the RMT behavior sets in after a Thouless time governed by diffusive charge transport. The $U(1)$ Floquet calculation required appending an unconstrained qudit to a qubit whose charge is conserved, resulting in an enlarged 
local Hilbert space $\mathbb{C}^2 \otimes \mathbb{C}^q$ of dimension $2q$. Once again, the limit $q\rightarrow \infty$ was necessary to make analytic predictions although, just like the case without $U(1)$ symmetry, the results were expected to hold more generally i.e. even in the case of $q=1$. 

In this work we consider thermalization dynamics in various $U(1)$ conserving random Floquet circuits. Surprisingly, we find that the system deviates from robust thermalization for the minimal Haar random $U(1)$ conserving Floquet model with nearest neighbor gates, spin 1/2 qubits, and no additional qudits (i.e. with $q=1$). While the random-in-time Haar random $U(1)$ conserving circuit with $q=1$ is  provably diffusive~\cite{Khemani_2018, Rakovszky_2018}---and the same behavior was expected to occur in Floquet circuits~\cite{Friedman_2019}---we instead find subdiffusive behavior of conserved charges for long times and slow thermalization dynamics for numerically accessible system sizes. We compute various local and global metrics of quantum thermalization, such as level spacing statistics, the transport of conserved charges, and the dynamics of entanglement. While all metrics show trends that are consistent with asymptotic thermalization for large enough system sizes and times, numerically accessible sizes and times do not show robust thermalization. By studying the parameterization of the gates, we argue that the slow dynamics can be understood in terms of proximity of the nearest neighbor Haar random $U(1)$ conserving gates to localized regimes in parameter space. 

In contrast, robust thermalization is recovered by taking $q>1$ or, analogously, by increasing the range of gates $r$, or the period $T$. Random circuits (with and without symmetries) have come to serve as canonical models for numerically and experimentally studying quantum dynamics. Our work informs future studies of charge conserving Floquet circuits by showing that the minimal such circuit is \emph{not} robustly chaotic, and needs to be extended in one of the ways mentioned above, i.e. by increasing the range, the period, or $q$, to achieve robust thermalization.  

More generally, our work shows that the combination of locality, symmetry, and a small local Hilbert space can quite generically hinder thermalization and exploration of the (constrained) Hilbert space. Interestingly, it was recently pointed out \cite{Marvian_2022}  that local symmetric gates cannot produce the full group of global symmetric gates even for random-in-time systems (regardless of the range of interactions). 
This is to be contrasted with the fundamental result in quantum computing that any global unitary in $U(N)$ can be arbitrarily approximated by a sequence of 2-local gates \cite{PhysRevA.51.1015,PhysRevLett.75.346}. Whereas the result in \cite{Marvian_2022} does not preclude the emergence of local thermalization, it raises intriguing questions of how truly random the global unitaries generated by local and symmetric evolution are.

The rest of this paper is organized as follows. We begin in Section \ref{sec:model} with a description of the models considered in this work. We show how to parameterize two-site Haar random $U(1)$ gates in two different ways: By directly constraining transition amplitudes in the unitary gate in accordance with the symmetry, and in terms of the generators of the Lie algebra of $U(4)$ that commute with the generator of the $U(1)$ subgroup on two qubits. The various Lie algebra coordinates, which we refer to as "couplings", serve as important tuning knobs of the models, controlling the interaction strength and real and complex hopping amplitudes.  We find that the minimal nearest neighbor $U(1)$ conserving Floquet circuit fails to saturate to the RMT prediction at numerically accessible system sizes, and explain this using the parameterization of $U(1)$ gates. We present level statistics data in Section \ref{sec:ratiofactor},  charge transport in Section \ref{sec:charge_transport} and  entanglement dynamics in Section \ref{sec:entanglement}. In all cases, the minimal slowly thermalizing model is contrasted with a robustly thermalizing model with larger period $T>1$, longer range interactions $r>1$, or enlarged local Hilbert space $q>1$. We summarize and conclude in Section \ref{sec:discussion}.

\section{Models \label{sec:model}}

\subsection{Circuit Architecture}

In this subsection, we introduce the general architecture of $U(1)$ conserving Floquet circuits considered in this work. The circuit structure will be labeled by three parameters: the local Hilbert space dimension $2q$, the range $r$, and the period $T$, discussed in turn below. The unitary gates comprising the circuit are drawn from various probability distributions, which are discussed in the next subsections.

We study periodic one-dimensional chains of length $L$ sites, where the degree of freedom on each site is the direct product of a spin-1/2 qubit and a qudit with Hilbert
space dimension $q$. Thus, the local Hilbert space is $\mathbb{C}^2 \otimes \mathbb{C}^{q}$, with dimension $2q$. 
We can label the spin state on each site $i$ as $(\uparrow a)_i$ or $(\downarrow b)_i$, where the first label is the spin state in the Pauli $z$ basis and the second label is the qudit state. Likewise, $X_j,Y_j,Z_j$ denote Pauli matrices acting on the qubit on site $j$, where the identity operator acting on the qudit state is left implicit. The time-evolution is constrained to conserve the total $z$ component of the spin-1/2’s, $S_z^{\rm tot} = \sum_j Z_j$, while the qudits are not subject to any conservation laws~\cite{Khemani_2018}. 

We consider brickwork circuits with staggered layers of range-$r$ unitary gates. The range $r$ is the number of contiguous qudits each individual gate acts on, so that $U_{j, j+1, \cdots ,j+(r-1)}$ acts on sites $(j, j+1, \cdots ,j+r-1)$. Thus, $r=2$ denotes nearest-neighbour gates while $r=3$ is a three-site gate including both nearest and next-nearest neighbour interactions. Each gate is a $(2q)^r \times (2q)^r$ block diagonal matrix, with $(r+1)$ blocks labeled by the total $z$ charge of the qubits on $r$ sites: $\sum_{i=j}^{j+r-1} Z_i$.  The blocks have size $\binom{r}{n} q^r$ with $n=0,1,...,r$.  For example, for $r=2$, the gates $U_{j, j+1}$ have the structure:  
\begin{equation}
    U_{j,j+1} = {\tiny\begin{pmatrix} 
    \uparrow \uparrow & &  \\
    \begin{pmatrix}  \\  \\  q^2 \times q^2 \\ \\  \\  \end{pmatrix}  \\ 
    & \uparrow\downarrow, \downarrow\uparrow & \\
    & \begin{pmatrix} & & \\  & & \\  & & \\  & 2q^2 \times 2q^2 & \\   & &\\  & & \\  & & \\\end{pmatrix} & \\
    % & \begin{pmatrix} . &.\\.&.\end{pmatrix} &  \\ 
    & & \downarrow \downarrow \\
    & &  \begin{pmatrix}  \\  \\  q^2 \times q^2 \\ \\  \\  \end{pmatrix}   \end{pmatrix}},
    \label{eq:U1unitary2site}
\end{equation}
comprised of  (i) a $(q^2 \times q^2)$ block acting in the $(\uparrow a)_i \otimes (\uparrow b)_{i+1}$ subspace, (ii) a $(2q^2 \times 2q^2)$ block acting in the $(\uparrow a)_i \otimes (\downarrow b)_{i+1}, (\downarrow a)_i \otimes (\uparrow b)_{i+1}$ subspace, and (iii) a $(q^2 \times q^2)$ block acting in the $(\downarrow a)_i \otimes (\downarrow b)_{i+1}$ subspace. Likewise, for $r=3$, the gates $U_{j, j+1, j+2}$ have the structure:  
\begin{align}
&U_{j,j+1, j+2} =\nonumber \\
&{\tiny 
    \begin{pmatrix} 
    \uparrow \uparrow \uparrow & & & \\
    \begin{pmatrix}    \\ \\ q^2 \times q^2 \\ \\ \\   \end{pmatrix}  & & &  \\ 
    & \uparrow \uparrow \downarrow, \uparrow \downarrow\uparrow, \downarrow \uparrow\uparrow & & \\
    & \begin{pmatrix} & & \\  & & \\  & & \\  & 3q^2 \times 3q^2 & \\   & &  \\  & & \\ & & \\ \end{pmatrix} & & \\
    & &\downarrow \downarrow \uparrow, \downarrow \uparrow\downarrow, \uparrow \downarrow\downarrow & \\
    & &\begin{pmatrix}  & & \\ & & \\  & & \\  & 3q^2 \times 3q^2 & \\   & &  \\  & & \\ & & \\\end{pmatrix} &  \\
    & & &\downarrow \downarrow \downarrow\\
    & &  & \begin{pmatrix}    \\ \\ q^2 \times q^2 \\ \\ \\    \end{pmatrix}   \end{pmatrix}},
    % =\begin{pmatrix} (.) & & & \\ & \begin{pmatrix} . &.&.\\.&.&.\\ .&.&.\end{pmatrix} \\ & & \begin{pmatrix} . &.&.\\.&.&.\\ .&.&.\end{pmatrix} & \\ & & & (.)\end{pmatrix}.
    \label{eq:U1unitary3site}
\end{align}
comprised of  (i) a $(q^2 \times q^2)$ block acting in the $(\uparrow a)_i \otimes (\uparrow b)_{i+1} \otimes (\uparrow c)_{i+2}$ subspace, (ii) a $(3q^2 \times 3q^2)$ block acting in the $(\uparrow a)_i \otimes (\uparrow b)_{i+1} \otimes (\downarrow c)_{i+2},
(\uparrow a)_i \otimes (\downarrow b)_{i+1} \otimes (\uparrow c)_{i+2}, (\downarrow a)_i \otimes (\uparrow b)_{i+1} \otimes (\uparrow c)_{i+2}$  subspace,  (iii) a $(3q^2 \times 3q^2)$ block acting in the $(\downarrow a)_i \otimes (\downarrow b)_{i+1} \otimes (\uparrow c)_{i+2}, (\downarrow a)_i \otimes (\uparrow b)_{i+1} \otimes (\downarrow c)_{i+2}, (\uparrow a)_i \otimes (\downarrow b)_{i+1} \otimes (\downarrow c)_{i+2}$  subspace, and (iv) a $(q^2 \times q^2)$ block acting in the $(\downarrow a)_i \otimes (\downarrow b)_{i+1} \otimes (\downarrow c)_{i+2}$ subspace. 

\begin{figure}[]
\centering
\includegraphics[width=.48\textwidth]{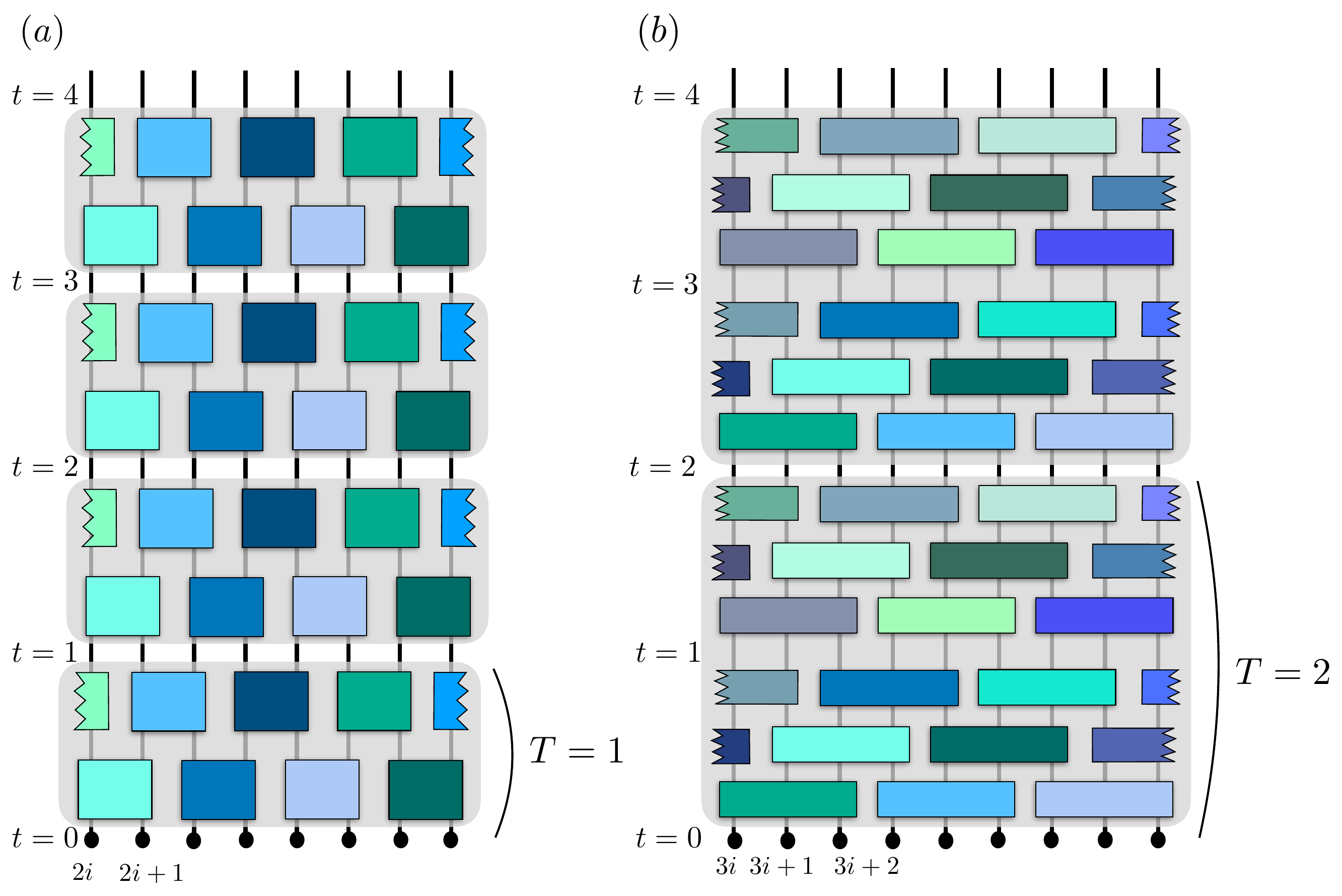}
\caption{(a) Brickwork circuit for the minimal model with nearest-neighbor gates with $r=2$, and Floquet period $T=1$, and (b) Circuit architecture with range $r=3$ gates and Floquet period $T=2$. Each gate locally conserves $S^z_{\rm tot}$.}
\label{fig:circuit}
\end{figure}

The circuit architecture has a periodic brickwork layout with variable period $T \in \mathbb{Z}$. The circuit implements discrete time evolution, and advancing by one unit of time comprises the application of a ``layer'' comprised of $r$ staggered sub-layers. Each sub-layer is displaced by one lattice site with respect to the prior sublayer (See Fig.~\ref{fig:circuit}). For example, for $r=2$, advancing by one unit of time entails applying one layer of even and odd gates: 
\begin{equation}
    U(t+1,t) = \underbrace{\prod_j U_{2j+1, 2j+2}(t)}_{U_{\rm odd}(t) \equiv U_1(t)}  \underbrace{\prod_i U_{2i, 2i+1}(t)}_{U_{\rm even}(t) \equiv U_0(t)} 
\end{equation}
Likewise, $r=3$ requires applying three staggered sub-layers of gates starting from the $(0,1,2)$, $(1,2,3)$ and $(2,3,4)$ bonds respectively:
\begin{align}
    U(t+1,t) = &\underbrace{\prod_k U_{3k+2, 3k+3, 3k+4}(t)}_{U_2(t)} \times \nonumber \\  
    &\underbrace{\prod_j U_{3j+1, 3j+2, 3j+3}(t)}_{ U_1(t)} \times \nonumber \\ 
    &\underbrace{\prod_i U_{3i, 3i+1, 3i+2}(t)}_{U_0(t)} 
\end{align}
More generally, for range $r$ gates, 
\begin{align}
     U(t+1,t) &= \prod_{\alpha = 0}^{r-1} U_\alpha(t) \nonumber \\
     U_\alpha(t) &= \prod_j U_{rj + \alpha, rj + \alpha+ 1, \cdots, rj + \alpha+ r-1 }
\end{align}

For a circuit with periodicity $T$, the gates in the first $T$ layers are chosen independently,  and layers repeat after $T$ time-steps: $U(t+T+1,t+T) = U(t+1,t)$. The Floquet unitary is defined as the time-evolution operator for period $T$:
\begin{equation}
    U_F(T) = \prod_{t=0}^{T-1} U(t+1,t),
\end{equation}
and $U(t=nT, 0) = U_F(T)^n$. 

\subsection{Haar Random U(1) Conserving Gates}

A central focus of our study is on the most random $U(1)$ Floquet dynamics, where each symmetry block in each unitary gate in the first $T$ layers is sampled randomly and independently from the Haar distribution. This requires uniformly sampling unitaries in $U(N)$ for all the $(r+1)$ block sizes $N=\binom{r}{n} q^r, n = 0,1\cdots,r$ in each independent gate which can be done, for instance, with the numerical sampling algorithm described in \cite{ozols}. For example, $r=2$ entails  sampling 3 independent blocks for each independent gate, of sizes $N= q^2, 2q^2$ and $q^2$ [cf. Eq.~\eqref{eq:U1unitary2site}]. Similarly,   $r=3$ requires sampling 4 independent blocks for each independent gate, of sizes $N= q^2, 3q^2, 3q^2$ and $q^2$ [cf. Eq.~\eqref{eq:U1unitary3site}].

In what follows, we will show that the \emph{minimal} Haar random U(1) conserving Floquet model with nearest neighbor dynamics on spin 1/2 qubits (no qudits), $q=1$, $r=2$, $T=1$, is \emph{not} robustly thermalizing for the accessible system sizes. In contrast, the simplest extensions of the minimal model, $q=2$, $r=3$, or $T=2$, all show robust thermalization (cf. Section~\ref{sec:ratiofactor}).  All these extensions have the effect of making the Floquet dynamics less locally constrained, which aids with thermalization.

The balance of the paper is almost entirely focused on dynamics with $r=2, q=1$, and we consider both $T=1,2$. The $T=2$ model will serve as a strongly chaotic reference model to be contrasted with the minimal model. To understand the origin of  slow thermalization in the minimal model, we will find it useful to consider deformations of the Haar random gates that illuminate the proximity of nearby localized regimes in parameter space. To this end, we will find it useful to explicitly parameterize two-site $U(1)$ conserving gates in multiple ways, which we turn to next.

\subsection{Parameterizing two-site $U(1)$ conserving gates}\label{sec:parametrize_minimal_model}

The unitary gates for the minimal model with $r=2, q=1$ can be parameterized explicitly. The two-site basis is just $\{\ket{\uparrow\uparrow},\ket{\uparrow\downarrow},\ket{\downarrow\uparrow},\ket{\downarrow\downarrow}\}$, and $U_{j,j+1}$ takes the form 

\begin{equation}
    U_{j, j+1} = \begin{pmatrix} e^{i\alpha} & & \\ & e^{i\beta}  \begin{pmatrix} \cos(\psi) e^{i\eta} & \sin(\psi) e^{i\chi} \\ - \sin(\psi) e^{-i\chi} & \cos(\psi) e^{-i\eta} \end{pmatrix} \\ & & & e^{i\gamma}\end{pmatrix}.
    \label{U1unitary}
\end{equation}

The states $\ket{\uparrow \uparrow }$ and $\ket{\downarrow\downarrow}$ can only acquire a  phase, whereas the central block enacts a general unitary transformation on the $\{\ket{\uparrow \downarrow},\ket{\uparrow \downarrow}\}$ subspace. The gate is parameterized with six real parameters $\{\alpha, \beta, \gamma, \psi, \phi,\chi\}$. We will consider different distributions of these parameters in \ref{sec:Haar},\ref{sec:perturbedAnderson},\ref{sec:diagonal}.

We will also find it useful to parameterize the two-site unitary gate in terms of the generators $\{H^\alpha\}$ of the Lie algebra of $U(4)$ that commute with the generators of the $U(1)$ subgroup on two qubits. The local unitary gate in Eq.~\eqref{U1unitary} has six degrees of freedom, thus the Lie algebra contains six generators:
\begin{align}
U_{j, j+1}&= e^{i \sum_{\alpha=0}^5 c_\alpha H_{j,j+1}^\alpha}. \label{U1generators}
\end{align}
The choice of generators is not unique, but a natural basis is
\begin{align}
\{ H^\alpha_{j,j+1}\}=&\{I_j I_{j+1},\notag \\ &(Z_j+Z_{j+1})/2, \notag \\
    &Z_jZ_{j+1}, \notag \\
    & (X_jX_{j+1}+Y_jY_{j+1})/2, \notag \\
    & i(X_jY_{j+1}-Y_jX_{j+1})/2, \notag \\
    & (Z_{j+1}-Z_j)/2\}, \label{eq:U1generatorsbasis}
\end{align}
where $I$ is the identity matrix. In the chosen basis \eqref{eq:U1generatorsbasis}, the first three generators commute with all other generators, $[H_{0,1,2},H_\alpha]=0, \, \> \forall \> \alpha$, therefore we can factorize the unitary as
\begin{align}
U_{j, j+1}&=\big( \prod_{\alpha=0}^2 e^{i c_\alpha H^\alpha_{j,j+1}} \big)e^{i\sum_{\alpha=3}^5 c_\alpha H^\alpha_{j,j+1}}
\label{eq:U1generators_factor}
\end{align}

In order to make connections to the physics of localization, we express the spin operators in Eq.~(\ref{eq:U1generatorsbasis}) in terms of fermionic operators via a Jordan Wigner transformation. We denote $\hat{f}_{j}$ as fermionic operators $\{\hat{f}_{j},\hat{f}_{j}^\dagger\} = 1$ and $\hat{n}_{j} = \hat{f}_{j}^\dagger\hat{f}_{j} = (I_j-Z_{j})/2$ are density operators. It is straightforward to show that the basis of generators in Eq.~(\ref{eq:U1generatorsbasis}) can be written as:
\begin{align}
    \{ H^\alpha_{j,j+1}\}=&\{I_j I_{j+1},\notag \\ &I_jI_{j+1}-\hat{n}_{j}-\hat{n}_{j+1}, \notag \\
    &I_jI_{j+1}-2(\hat{n}_{j}+\hat{n}_{j+1})+4\hat{n}_{j}\hat{n}_{j+1}, \notag \\
    &\hat{f}_{j}^{\dagger} \hat{f}_{j+1} + \hat{f}_{j+1}^{\dagger}\hat{f}_{j},\notag \\ 
    & i(\hat{f}_{j}^{\dagger}\hat{f}_{j+1}-\hat{f}_{j+1}^{\dagger}\hat{f}_{j}),\notag \\ 
    &\hat{n}_{j}-\hat{n}_{j+1} \}. \label{eq:fermion}
\end{align}
We can now interpret the physical meaning of each of the couplings $c_j$: $c_0$, $c_1$ and $c_5$ control the strength of the local potential, $c_3$ and $c_4$ control the (complex) nearest-neighbor fermion hopping, and $c_2$ controls the strength of the density-density interaction.

If we impose $c_2 = 0$, then the gates are non-interacting, yielding a disordered free fermion Floquet model which is known to be Anderson localized in 1D.  Interactions, on the other hand, tend to delocalize the system. 
Importantly, due to the Floquet nature of the model, interactions cannot be made arbitrarily large. In particular, the parameter $c_2$ is periodic modulo $2\pi$. Likewise, if we set $c_3, c_4$ to be zero, this turns off the hopping, giving a model which is diagonal in the $z$ basis and is trivially localized. 

As we will see, although diagnostics of chaos suggest that the minimal circuits with $T=1$ asymptotically thermalize, the proximity of these localized regimes and the limited number of parameters leads to slow thermalization at accessible sizes. In contrast, the $T=2$ circuits are less constrained in parameter space and show robust thermalization. We now discuss three families of two-site $U(1)$ gates we consider in this work, obtained by varying the parameters $\{c_\alpha\}$. For each family, we examine $T=1$ as a minimal model and $T=2$ as a reference thermalizing model.

\subsubsection{Haar random $U(1)$ gates}
\label{sec:Haar}
Our main results pertain to Haar random $U(1)$ conserving gates and perturbations thereof. To study these perturbations, we will find it useful to present the Haar random gates in terms of the two parameterizations just introduced: In terms of the representation of $U(1)$ on two qubits \eqref{U1unitary}, and in terms of the corresponding Lie algebra in \eqref{U1generators}. The former is straight-forward, each block in ~\eqref{U1unitary} is sampled independently from the uniform Haar measure. In terms of the chosen parameters, this means that the three diagonal phases $(e^{i\alpha},e^{i\beta},e^{i\gamma})$ are all elements of $U(1)$, hence we can sample $(\alpha,\beta,\gamma)$ uniformly in $[0,2\pi]$. The central block (modulo the overall phase) parameterizes $SU(2)$. To sample from this manifold uniformly, we take $t=[\sin(\psi)]^2$ uniformly in $[0,1]$ and $(\eta,\chi)$ uniformly in $[0,2\pi]$. 

We turn to finding the equivalent distributions in terms of the couplings $(c_\alpha)$ of \eqref{U1generators}. As before, we can sample $(c_0,c_1,c_2)$ uniformly in $[0,2\pi]$ since the couplings $(c_0,c_1,c_2)$ are associated with the commuting part of \eqref{eq:U1generators_factor}. However, the remaining three generators $H^3=(XX+YY)/2, H^4=i(XY-YX)/2,$ and $H^5=(IZ-ZI)/2$ do not commute, and, consequently, the probability distributions of associated couplings $(c_3,c_4,c_5)$ do not factorize. Thus we have to find the volume element directly in terms of the Lie algebra~\footnote{Naively, one could try to find those distributions by a change of variables between $(\psi,\eta,\chi)$ in \eqref{U1unitary} and $(c_3,c_4,c_5)$ in \eqref{U1generators}, but the exponential map between \eqref{U1unitary} and \eqref{U1generators} does not allow for explicit solving of the new variables.}. Since we've already factored out the commuting part, the remaining algebra is that of $SU(2).$ To see this explicitly, the generators $H^{3,4,5}$ act only within the central block, and by defining effective spins $\ket{0}\equiv \ket{\downarrow \uparrow}, \ket{1}\equiv\ket{\uparrow \downarrow}$, they act just like the single-site Paulis $(X,Y,Z)$, which obey the $SU(2)$ algebra. 
The details of the calculation are carried out in Appendix~\ref{sec:appendixA}, where we first discuss how to generically obtain the uniform (Haar) measure on a Lie group, and then specify to the $SU(2)$ case. Changing to polar coordinates, $c_3=R\sin(\theta) \cos(\phi)$, $c_4=R\sin(\theta) \sin(\phi)$, and  $c_5= R\cos(\theta)$, the distributions we find are $P(R)dR= \frac{2}{\pi} \sin(R)^2 dR,\>\> R \in [0,\pi], P(\theta)d\theta = \frac{1}{2} \sin(\theta) d\theta,\>\> \theta \in[0,\pi]$ and $P(\phi)d\phi = \frac{1}{2\pi} d\phi, \>\> \phi \in [0,2\pi]$. Sampling from these distributions, and substituting  the coordinate change gives us the desired distributions on $(c_3,c_4,c_5)$. Importantly, note that we cannot write down independent distributions for $(c_3,c_4,c_5)$ since the couplings are correlated. 

\subsubsection{Perturbed Anderson Localized Model}
\label{sec:perturbedAnderson}

As shown in Eq.~\eqref{eq:fermion}, the strength of interactions is controlled by $c_2$, with $c_2=0$ being a non-interacting Anderson localized model. We define a family of gates in which $c_2$ is non-random and tuned continuously from $0$ to $2\pi$, while the remaining 5 parameters are drawn using the same distributions as for the Haar random case described in the previous subsection. The ``interaction unitary'' can be expanded as $e^{ic_2 ZZ}=\cos(c_2)II +i \sin(c_2) ZZ$. For 
$c_2=\{0,\pi/2\}$, this reduces to $\{II,ZZ\}$, which results in a free theory. Note that the $\pi/2$ point is free because the $ZZ$ gates from the even and odd layers combine to give the identity. The middle value $c_2=\pi/4$, reduces to the symmetric superposition, $(II+iZZ)/\sqrt{2}$, corresponding to the most ergodic point. Meanwhile, the behavior for $c_2= \{0\pm \delta, \pi/2 \pm \delta\}$ is the same up to signs of the coefficients of $II,ZZ$, making the phase diagrams reflection symmetric about  $\pi/2$ and $\pi$.

\subsubsection{Perturbed Diagonal Model}
\label{sec:diagonal}

Next, we tune the real and complex hopping strength parameterized by $c_3$ and $c_4$. In the absence of hopping, the model is trivially localized since all other terms are diagonal in the $z$ basis. 
As discussed in \ref{sec:Haar}, the values $c_3,c_4$ are correlated with $c_5$, as the generators $H^{3,4,5}$ are non-commuting. For simplicity, we set $c_5=0$. This should qualitatively not affect the statistical properties of $U_{j,j+1}$: the generator $H^5$ produces a local potential, but so do $H^0$ and $H^1$. Even if $c_5=0$, the composition of even and odd gates still results in a potential that is random in space. This leaves us with $c_3=R\cos(\phi)$ and $c_4=R\sin(\phi)$. In order to have a single tuning parameter, we sample $\phi$ as prescribed by the Haar distribution, and tune a non-random $R$ from $0$ to $1$, set to be equal for all $U_{j,j+1}$. 

\begin{figure}[h!]
\centering
\includegraphics[width=.5\textwidth]{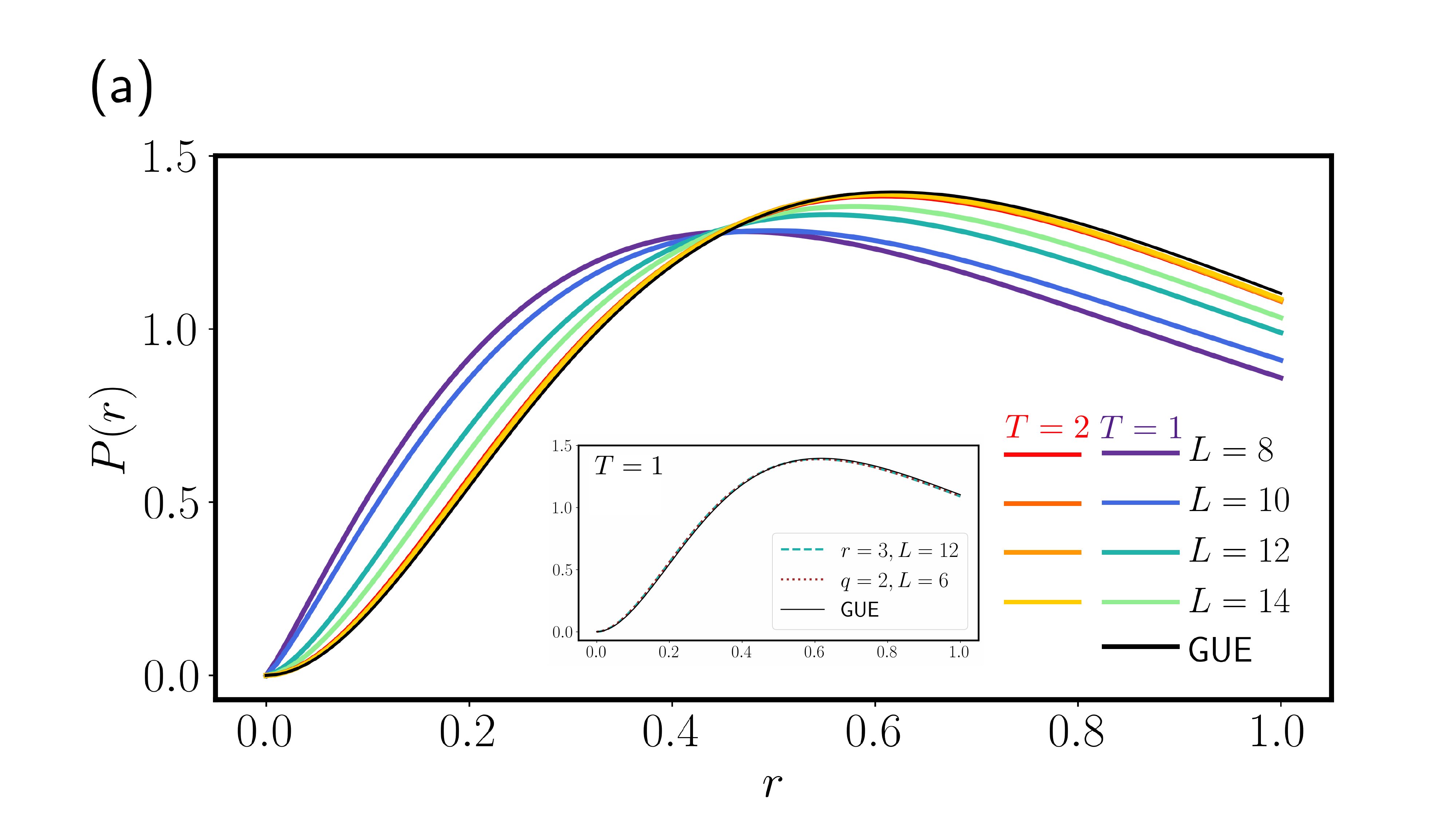}
\includegraphics[width=.45\textwidth]{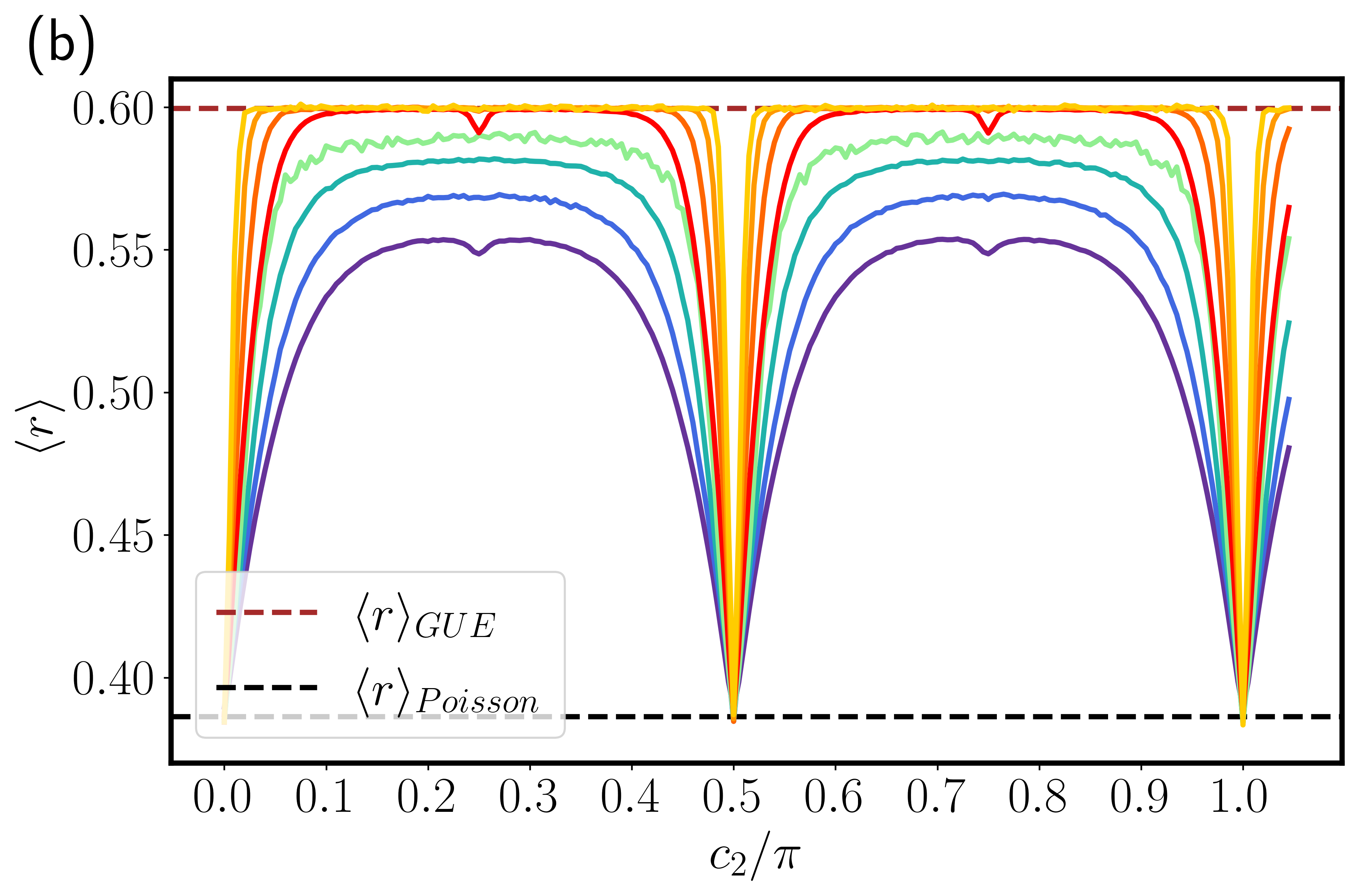}\\
\includegraphics[width=.45\textwidth]{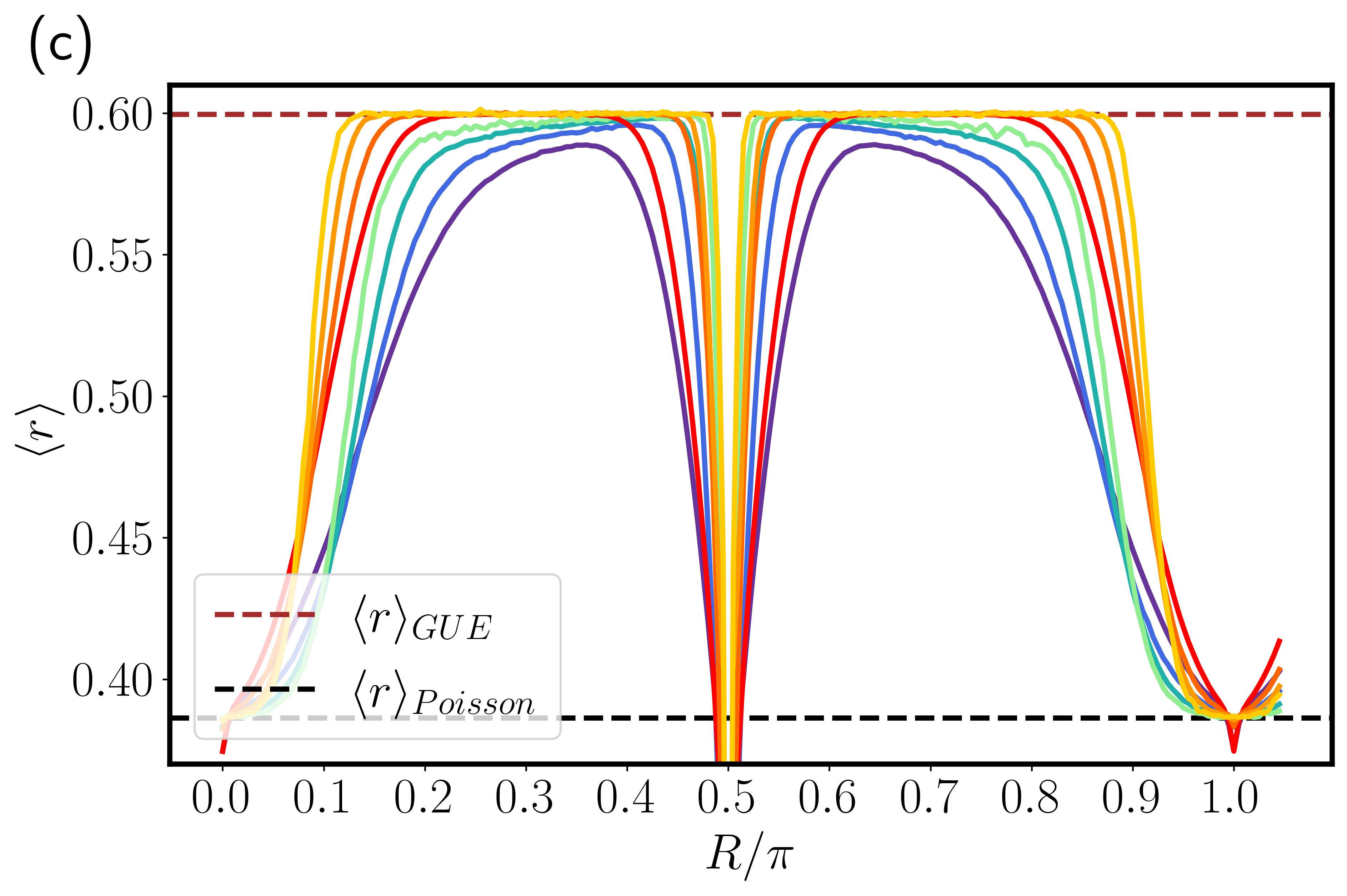}\\
\caption{(a) Distribution of the ratio factor for $T=1$  (blue color palette) and $T=2$ (red color palette) Haar random $U(1)$ conserving Floquet circuits as function of system size ($r=2$, $q=1$). Although the finite-size scaling indicates that the $T=1$ Haar random $U(1)$ circuit converges to the GUE distribution, this convergence is slow and numerically accessible system sizes are not robustly thermal. In contrast, the $T=2$ circuit exhibits GUE statistics even for the smallest system size. (inset): Extensions of the minimal model with $q=2$ and $r=3$. Both extensions display GUE statistics even for small sizes and $T=1$.  (b-c) Finite-size scaling of $\langle r \rangle$ as a function of (b) the interaction strength $c_2$ in the perturbed Anderson localized model and (c) the hopping strength $c_3 = R \cos(\phi)$, $c_4 = R \sin(\phi)$ in the perturbed diagonal model.  See Eqs.(\ref{U1generators})-(\ref{eq:U1generators_factor}) for parameter definitions and Sections \ref{sec:perturbedAnderson} and \ref{sec:diagonal} for model definitions. Once again, the $T=2$ circuit is much more robustly thermal. Number of disorder realizations for $L=8,10,12,14$ are $10^5, 10^4, 10^4, 10^3$ respectively.}
\label{fig:ratio_factor_c2_fixed_R_fixed_phisampled}
\end{figure}

\section{Level statistics} \label{sec:ratiofactor}

As a first diagnostic of thermalization, we study the
statistics of the eigenspectrum of the Floquet unitaries $U_F$. The eigenvalues are phases $e^{i\theta_n}$, which define quasienergies $\theta_n$ modulo $2\pi$. We compute the ratio of quasienergy spacings \cite{Oganesyan_2007}:
\begin{align}
    r_n &= \frac{ \text{min}(s_n,s_{n-1})}{ \text{max}(s_n,s_{n-1})} \label{r},
    %\>\> \langle r\rangle = \frac{1}{N} \sum_{n=1}^N r_n,
\end{align}
which captures level repulsion, a characteristic signature of random matrix behavior~\cite{2004Mehta_RMT}. In \eqref{r}, $s_{n}=\theta_{n+1}-\theta_n$ is the spacing between the $n, (n+1)$st quasienergies in sorted order. We recall that one needs to resolve all symmetries for level spacings to be an accurate diagnostic of chaos. In our case, we consider the $S_{\rm tot}^z = 0$ symmetry sector and look at the distribution of the $r$-ratio across the spectrum and across circuit realizations. We study both the full probability distribution $P(r)$ of the ratio, and  its mean  $\langle . \rangle$ (Fig.~\ref{fig:ratio_factor_c2_fixed_R_fixed_phisampled}). Thermalizing systems exhibit level repulsion akin to random matrix theory, hence we compare $P(r)$ and $\langle r\rangle$ to the predictions of the Gaussian Unitary Ensemble (GUE), with $P(r)_{\rm GUE} \sim \frac{(r+r^2)^2}{(1+r+r^2)^{1+3/2}}$ (ignoring the normalization constant), and $\langle r \rangle_{\rm GUE} \approx 0.60$ \cite{Atas_2013}. In contrast, integrable models exhibit uncorrelated level statistics and a Poisson distribution $P(r)_{P} \sim e^{-r}$ and $\langle r \rangle_P = 0.39$ \cite{1977PRSL_Berry}. 

 Fig.~\ref{fig:ratio_factor_c2_fixed_R_fixed_phisampled}(a) shows the distributions $P(r)$ for $T=1$ (blue colors) and $T=2$ (red colors) $U(1)$ Haar random Floquet circuits with $q=1, r=2$ defined in \ref{sec:Haar}. We observe that, despite the small difference between these circuits, the $T=1$ circuit approaches the GUE distribution significantly slower than the $T=2$ circuit, which matches the GUE distribution even for moderately small system size $L=8$. The inset shows data extending the minimal model either by adding a qudit with $q=2$, or by increasing the range of the gates to $r=3$. Both extensions lead to robust convergence to the GUE distribution even at small sizes. 

To more systematically examine the terms that hinder thermalization in the minimal model, we analyze the perturbed Anderson localized and perturbed diagonal models discussed in Sections \ref{sec:perturbedAnderson}, \ref{sec:diagonal}. In the first case, we set $c_2$ in Eq.(\ref{eq:U1generators_factor}) to be equal for all unitaries $U_{j,j+1}$. Figure \ref{fig:ratio_factor_c2_fixed_R_fixed_phisampled}(b) shows the ratio factor as a function of $c_2$. As expected, the ratio factor is Poisson $\langle r\rangle_{P} \sim 0.39$ in the non-interacting regime $c_2 = 0$, and grows monotonically until the most ergodic point $c_2 = \pi/4$. Note that the maximum value of $\langle r \rangle$ for finite size $T=1$ circuits is well below the GUE prediction even at the most ergodic point. However, finite-size scaling suggests that $\langle r \rangle$ converges to the GUE predictions for all values $c_2\neq 0$ (modulo $\pi$) in the thermodynamic limit. In contrast, for the $T=2$ circuits we find that $\langle r \rangle$ rapidly approaches the GUE value, even for the smallest system size and relatively small values of $c_2$.

In the second case, we tune the strength of the hopping terms $(c_3,c_4)$, with $c_3=c_4=0$ corresponding to the classical disordered Ising model (Fig.~\ref{fig:ratio_factor_c2_fixed_R_fixed_phisampled}(c)). In the absence of hopping, the $T=1,2$ circuits are classical (diagonal) disordered spin models that do not agree either with the Poisson or GUE predictions. As hopping is turned on with a finite value of $R$, $\langle r \rangle$ slowly (rapidly) approaches the GUE predictions for $T=1$ ($T=2$) circuits.

Since the Haar random U(1) conserving gates sample from $c_2, R$ randomly, they approximately average over the range of behaviors observed in Fig.~\ref{fig:ratio_factor_c2_fixed_R_fixed_phisampled}(b,c), explaining the lack of robust thermalization in the $T=1$ model.

\begin{figure}[]
\centering
\includegraphics[width=.5\textwidth]{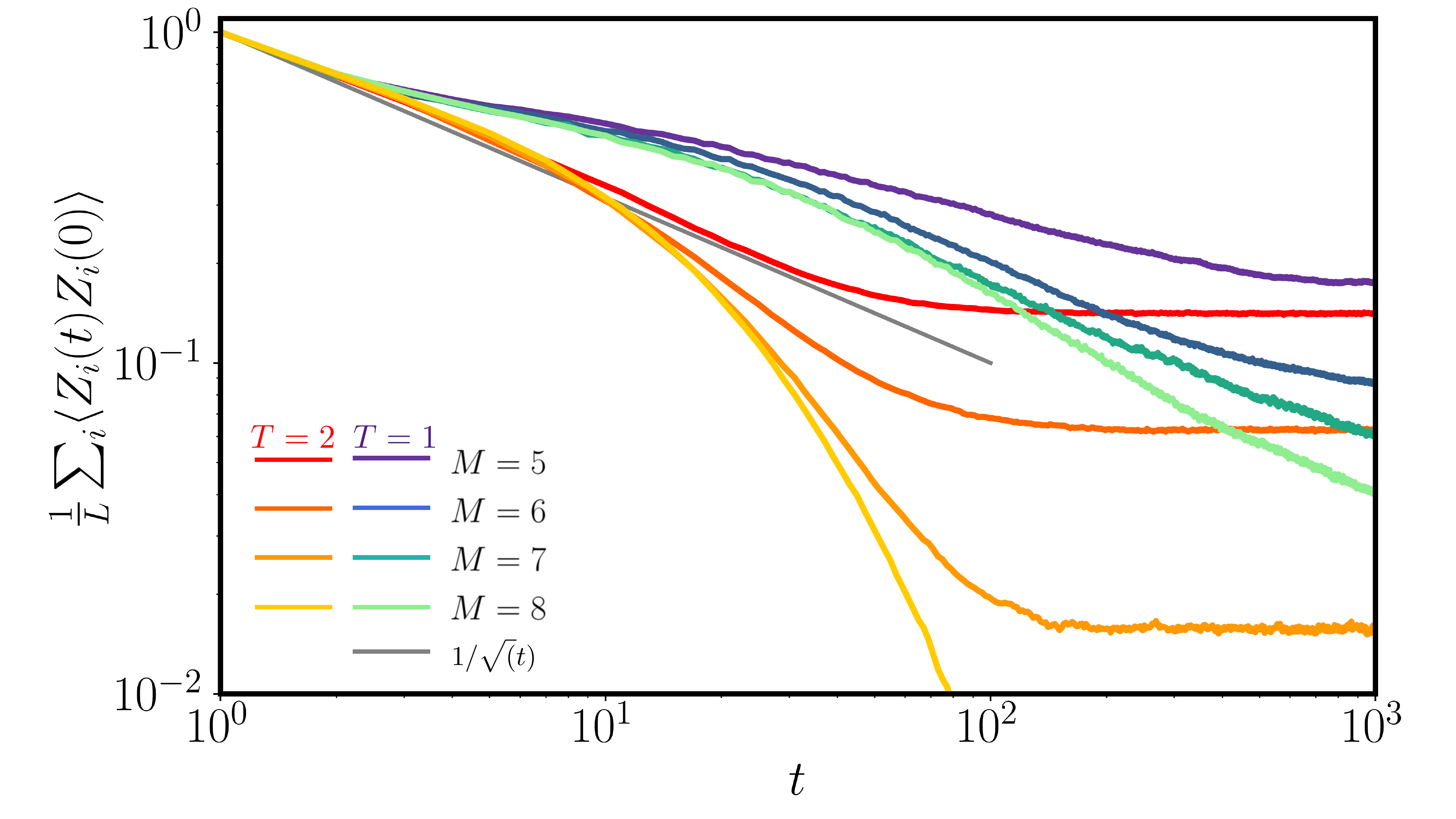}
\caption{  Spin-spin autocorrelation function for Haar random $U(1)$ Floquet circuits of size $L=16$, total spin $M = 5,6,7,8$,
%\textcolor{red}{$M=4,6,8,10$} 
averaged over $10^3$ circuit realizations. Data is shown for both the $T=1$ circuit  (blue color palette) and $T=2$ circuit (red color palette); both sets have $r=2, q=1$.  The solid gray line shows diffusive scaling $1/\sqrt{t}$. Whereas the $T=2$ circuits fits well within expectations from diffusive hydrodynamics, the $T=1$ circuits exhibits subdiffusive behavior for long times with a possible slow crossover towards diffusion.}

\label{fig:transport}
\end{figure}

\section{Dynamics of charge transport}\label{sec:charge_transport}

We now study thermalization of the local $U(1)$ charge density. For this purpose, we compute the spin autocorrelation function
\begin{equation}
    C(t) = \frac{1}{L}\sum_{i=1}^L \langle Z_i(t) Z_i(0) \rangle, 
\end{equation}
in states with fixed magnetization $\langle . \rangle = \langle \psi(M) |...|\psi(M)\rangle$. The state $\ket{\psi(M)}$ is chosen to have two domains so that all spins in the left domain are polarized $\uparrow$, all spins in the right domain are polarized $\downarrow$, and the relative sizes of the domains is set by $M$. The average over circuits is once again implicit in $C(t)$. For a chaotic system with only $U(1)$ symmetry, one expects charge diffusion across the system. The characteristic signature of diffusive modes in the autocorrelation function is the asymptotic dependence $C(t) \sim 1/\sqrt{t}$ at long times. Figure \ref{fig:transport} shows the autocorrelation function for the $T=1,2$ Haar random $U(1)$ Floquet circuits (blue and red color palette) for different values of total magnetization $M$ ($q=1, r=2$). Whereas $T=2$ circuits exhibit relaxation consistent with diffusion, we instead find subdiffusive relaxation of charge in $T=1$ circuits for a long period of time. While the data is consistent with a flow towards diffusion at later times, the slow relaxation of charge again points to the lack of robust thermalization in the $T=1$ circuit.

\section{Entanglement dynamics}\label{sec:entanglement}

\ignore{\begin{figure}[t!]
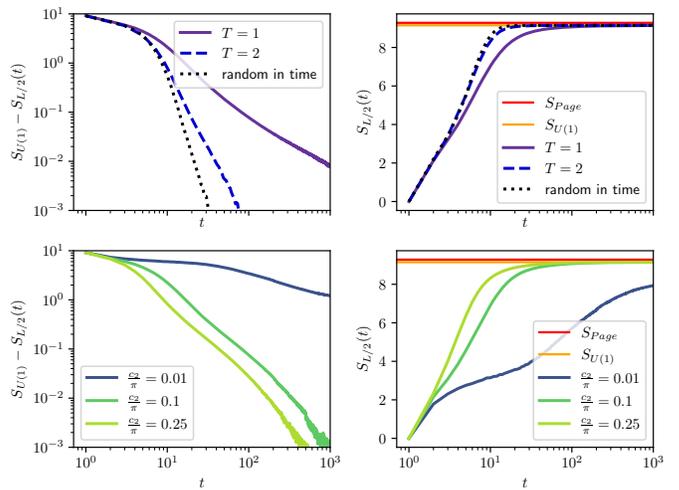

\centering
\includegraphics[width=.23\textwidth]{L20_SVN_U1_convergence.pdf}
\includegraphics[width=.23\textwidth]{L20_SVN_U1_growth.pdf}\\
\includegraphics[width=.23\textwidth]{L16_SVN_U1_c2_convergence.pdf}
\includegraphics[width=.23\textwidth]{L16_SVN_U1_c2_growth.pdf}
\caption{Evolution of the von Neumann entropy for half-system size $S_{L/2}(t)$. Top panels: $L=20$ averaged over 100 circuits for the $T=1$, $T=2$, and random in time Haar random $U(1)$ model defined in Section \ref{sec:Haar} ($q=1$, $r=2$). Bottom panels: $L=16$ averaged over 100 circuits for the perturbed Anderson localized model, with gate set parametrized by $c_2$ and described in Section \ref{sec:perturbedAnderson}. Panels (a) and (c) show convergence to the saturation value $S_{U(1)}$ of Haar random states with fixed magnetization $S^z_{tot}=0$ \cite{2019PRD_BianchiDona}, while (b) and (d) show entanglement growth with time (same data as in (a) and (c)). The Page value $S_{\rm Page}$ (the saturation value for Haar random states unconstrained by symmetry) is plotted as a reference \cite{Page_1993}. In all cases, the initial state is an antiferromagnetic state in the $S^z_{tot}=0$ sector.}
\label{fig:SVN}
\end{figure}}

\begin{figure}[t!]
\centering
\includegraphics[width=.5\textwidth]{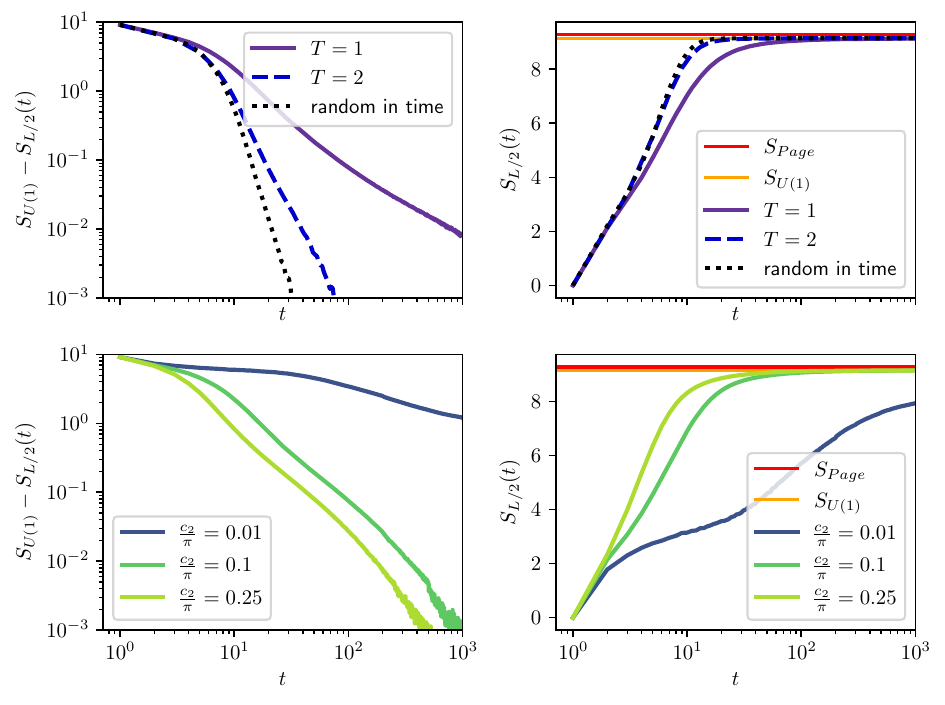}
\caption{Evolution of the von Neumann entropy for half-system size $S_{L/2}(t)$. Top panels: $L=20$ averaged over 100 circuits for the $T=1$, $T=2$, and random in time Haar random $U(1)$ model defined in Section \ref{sec:Haar} ($q=1$, $r=2$). Bottom panels: $L=16$ averaged over 100 circuits for the perturbed Anderson localized model, with gate set parametrized by $c_2$ and described in Section \ref{sec:perturbedAnderson}. Panels (a) and (c) show convergence to the saturation value $S_{U(1)}$ of Haar random states with fixed magnetization $S^z_{tot}=0$ \cite{2019PRD_BianchiDona}, while (b) and (d) show entanglement growth with time (same data as in (a) and (c)). The Page value $S_{\rm Page}$ (the saturation value for Haar random states unconstrained by symmetry) is plotted as a reference \cite{Page_1993}. In all cases, the initial state is an antiferromagnetic state in the $S^z_{tot}=0$ sector.}
\label{fig:SVN}
\end{figure}

We proceed to discuss the dynamics of entanglement, which is another metric of quantum thermalization in many-body systems. Various aspects of the evolution of entanglement, such as the rate of growth of entanglement as a function of time, depend on  the type of entanglement entropy being considered and the details of the initial conditions. As an example, for initial states with charge inhomogeneity evolving under charge conserving dynamics, the second Renyi entropy grows as $\sqrt{t}$ at early times, while the von Neumann entropy grows ballistically \cite{2019PRL_Rakovszky}.
In general,  we expect that the von Neumann entropy grows ballistically in local thermalizing systems and saturates to its maximum value in 
in ${\cal O}(L)$ time. 

Here we consider the evolution of the half-chain von Neumann entropy, $S_A = {\rm Tr}_A[\rho_A \log\rho_A]$, with $\rho_A = {\rm Tr}_B[|\Psi(t)\rangle\langle\Psi(t)|]$ obtained after tracing half of the degrees of freedom. We consider initial states in the $S^z_{tot}=0$ sector. To avoid  effects caused by slow relaxation of domain walls, we choose the antiferromagnetic initial state, $|\uparrow\downarrow\uparrow\downarrow\cdots\rangle$. 
Figure \ref{fig:SVN}(a)-(b) show the dynamics of entanglement for $T=1,2$ Haar random $U(1)$ Floquet circuits ($r=2, q=1$), as well as a random in time version. In panel (a), we measure the convergence to the saturation value of entanglement for Haar random states with $U(1)$ symmetry, denoted $S_{U(1)}$; this value is slightly different from the Page value for random states without symmetry restrictions~\cite{2019PRD_BianchiDona}. The direct comparison to the analytic prediction highlights the different growth behaviour more transparently than the simple entanglement growth picture in panel (b).

As expected, we find that the entanglement entropy relaxes to its saturation value in $\sim 20$ time steps for the random in time and $T=2$ circuit model, which is compatible with the time necessary for entanglement to spread across the entire system. On the other hand, the $T=1$ circuits do not reach their saturation value for timescales several orders of magnitude longer, and the approach to the final saturation value appears parametrically slower in time.

As before, to more systematically examine the terms controlling thermalization, we study entanglement growth for the $T=1$ perturbed Anderson localized circuit with gates parameterized by the integrability breaking term $c_2$ in Fig.~\ref{fig:SVN}(b). We observe that the strength of the interaction term $c_2$   controls the rate of entanglement growth. For all values of $c_2$, as well as $c_2$ sampled from the Haar distribution shown in panel (a), we find that the $T=1$ circuit takes at least an order of magnitude longer to thermalize as compared to the $T=2$ circuit, in agreement with the conclusions obtained from level statistics and the autocorrelation function.

\section{Summary}\label{sec:discussion}

We have studied thermalization in various $U(1)$ conserving random Floquet circuits. Surprisingly, we find the absence of robust thermalization for a minimal Haar random $U(1)$ conserving Floquet model with nearest-neighbor gates, spin 1/2 qubits, and periodicity $T=1$. While various diagnostics of chaos are consistent with asymptotic thermalization, numerically accessible system sizes show a slow approach to chaotic behavior, and subdiffusive dynamics for long times.  This behavior is surprising because the random-in-time version of the same model is provably diffusive~\cite{Khemani_2018, Rakovszky_2018} and, more generally, because localization is generally not expected in \emph{minimally structured} Haar random Floquet models in which neighboring spins are strongly coupled and there is no preferred local basis in which the disorder is strong~\cite{S_nderhauf_2018}.  We explain the slow thermalization in terms of proximity of the Haar random $U(1)$ conserving gates to localized regimes in parameter space by studying a convenient parameterization of the gates in terms of Lie algebra generators.  Our results imply that the interplay of symmetry, locality, and time-periodicity can constrain the dynamics enough to significantly slow down thermalization. 

Random Floquet models are a workhorse for numerically studying thermalization dynamics; our results should inform future numerical studies of charge conserving thermalizing Floquet circuits by updating the minimal $U(1)$ random Floquet model needed to obtain robust thermalization. Importantly, we show that increasing the Floquet period, the range of interactions, or the onsite Hilbert space dimension $q>1$, is sufficient to weaken constraints enough to show robust thermalization.

\section*{Acknowledgements}

We acknowledge insightful discussions with Chaitanya Murthy and Jon Sorce. This work was supported by the US Department of Energy, Office of Science, Basic Energy Sciences, under Early Career Award Nos. DE-SC0021111 (C.J and V.K.). J.F.R.-N. acknowledges support from the Gordon and Betty Moore Foundation’s EPiQS Initiative through Grant GBMF4302 and GBMF8686. V.K. also acknowledges support from the Alfred P. Sloan Foundation through a Sloan Research Fellowship and the Packard Foundation through a Packard Fellowship in Science and Engineering. We also acknowledge the hospitality of the Kavli Institute for Theoretical Physics at the University of California, Santa Barbara (supported by NSF Grant PHY-1748958).

%\bibliography{main}

%apsrev4-2.bst 2019-01-14 (MD) hand-edited version of apsrev4-1.bst
%Control: key (0)
%Control: author (8) initials jnrlst
%Control: editor formatted (1) identically to author
%Control: production of article title (0) allowed
%Control: page (0) single
%Control: year (1) truncated
%Control: production of eprint (0) enabled
%

\onecolumngrid
\appendix

\newpage

\section{Uniform volume element on $SU(2)$}\label{sec:appendixA}

In this appendix, we derive the volume element of a Lie Group $\mathcal{G}$ in terms of the coordinates of the Lie algebra generators. We then apply this to the example of $\mathcal{G}=SU(2)$. 

As a preface, we note that the more common way of defining the volume element is in terms of the fundamental representation $U$ of $\mathcal{G}$. For instance, in the case of $SU(2)$, we can easily derive the volume element by thinking of the group as a $3-$sphere embedded in $\mathbb{R}^4$, with $U$ parametrized as in the central block of \eqref{U1unitary}. 

Another approach is to parametrize $U$ in terms of Euler angles. In this case, $U$ is a \textit{product} of single qubit rotations along separate axes, $U=e^{ix X} e^{i y Y} e^{i z Z}$. In order to transform this into a rotation around a single axis, $U=e^{i(x X + y Y+ z Z)}$, one needs to employ the group composition law of $SU(2)$. 

A more straight-forward way to derive the volume element, which generalizes to any Lie group, is to do so directly in terms of the Lie algebra generators \cite{molinari}. The algebra of the generators specifies the metric tensor $g_{ab}$ in a given representation. Once this is known, we can carry out the group integral using the volume element 

\begin{align}
    dV = \sqrt{|\text{det}(g)|}, \label{dV}
\end{align}

which leads to the Haar measure. Let us start with Lie group $\mathcal{G}$. The representation of this group in terms of unitary matrices $U$ is obtained with the exponential map, 

\begin{align}
    U=e^{iH}, \>\> H(x)= \sum_{a=1}^n x_a T_a, 
\end{align}

where $n$ is the dimension of the real vector space spanned by the generators $\{T_a\}$ of the Lie algebra. We can always choose the generators to be orthogonal, $Tr(T_a T_b)= \delta_{ab}$, so that we can easily extract the coordinates $ x_a=Tr(T_a H)$. The algebra of the generators is encoded in the structure constants $f_{abc}$. 

\begin{align}
    [T_a,T_b] &= i \sum_c f_{abc} T_c \label{LieAlgbera}
\end{align}

The metric $g$ in a chosen parametrization is defined by the invariant measure

\begin{align}
    ds^2 = \sum_{ab} g_{ab}(x)dx_a dx_b = Tr(U^{-1}dU U^{-1}dU). 
\end{align}

$g$ is a real symmetric $n \times n$ matrix. The differential of a group element $dU$ is defined in \cite{derivative_exponential_map}, $dU = d(e^{iH}) =  \int_0^1 dt  e^{iH(1-t)} dH e^{iHt}$ . With this in hand, we can calculate the invariant measure,

\begin{align}
    ds^2 &= - \int_0^1 \hspace{-.3cm} dt \int_0^1 \hspace{-.3cm} ds Tr\big(e^{-iH} e^{i(1-t)H} dH e^{iHt} e^{-iH} e^{iH(1-s)} dH e^{iHs}\big)
    = \int_{-1}^1 \hspace{-.3cm}d\tau Tr \bigg( e^{-iH\tau} dH e^{iH\tau} dH \bigg) (1-|\tau|)
\end{align}

where we used cyclicity of the trace, and changed variables to a symmetrized version $w=\frac{1}{2}(s+t)$ and $\tau=t-s$, and also rescaled the integral. Since $dH=\sum_a dx_a T_a$, we can look at each term in $g_{ab}$ individually. 

\begin{align}
    g_{ab}&= \int_{-1}^1 d\tau Tr \bigg( e^{-iH\tau} T_a e^{iH\tau} T_b\bigg) (1-|\tau|) \label{gab}
\end{align}

We know that the $\{T_a\}$ form a basis, so let's write the time evolution of $T_a$ as an operator expansion,

\begin{align}
    e^{i\tau H(x)} T_a e^{-i \tau H(x)} &= \sum_b c_{ab}(\tau,x) T_b.
\end{align}

We extract $c_{ab}(\tau,x)$ using the orthonormality property of the generators, 

\begin{align}
    c_{ab}(\tau,x) = Tr\big(e^{i\tau H(x)} T_a e^{-i \tau H(x)} T_b\big) 
\end{align}

To solve for $c_{ab}(\tau,x)$, we write the differential equation

\begin{align}
    \frac{d}{d\tau} c_{ab}(\tau,x) &= - \sum_{c,d} x_c f_{cad} c_{db}(x,\tau) \equiv - M_{ad} c_{bd}(x,\tau),
\end{align}

where we've defined the matrix $M_{ab}= \sum_c x_c f_{cab}$ and used \eqref{LieAlgbera}. The solution is an exponential, $c_{ab}(\tau,x)=(e^{\tau M(x)})_{ab}$. Plugging this back into \eqref{gab}, we find

\begin{align}
    g_{ab}(x)&= \int_{-1}^1 d\tau (1-|\tau|) \big(e^{\tau M(x)}\big)_{ab}. 
\end{align}

Note that $M$ is real and anti-Hermitian. As such, its eigenvalues are purely imaginary, and the non-zero ones must come in pairs. Furthermore, the matrices $g$ and $M$ are diagonalized by the same unitary matrix, therefore, if $\lambda_i$ is an eigenvalue of $M$, the corresponding eigenvalue of $g$ is 

\begin{align}
    g_i &= \int_{-1}^1 d\tau(1-|\tau|) e^{i\tau \lambda_i} = \frac{\sin^2(\lambda_i/2)}{(\lambda_i/2)^2}
\end{align}

So far, this was applicable to any Lie group. In our case, we want to study the matrix $M$ corresponding to the algebra of $su(2)$. The matrix elements are given as $M_{ab}= \sum_{c} x_c f_{cab} = \sum_c x_c 2 \varepsilon_{cab}$, resulting in 

\begin{align}
    M&= \begin{pmatrix} 0 & - z & y \\ z& 0& -x\\ -y &x &0
    \end{pmatrix},
\end{align}

with eigenvalue equation $-\lambda^3 + \lambda (x^2+y^2+z^2) \equiv -\lambda^3 + \lambda R^2=0$, where we've defined $R=\sqrt{x^2+y^2+z^2}$. The resulting spectra are $\text{spec(M)}= \{0, 2R ,-2R\}$ and $\text{spec(g)} = \{1, \frac{\sin^2(R)}{R^2}, \frac{\sin^2(-R)}{R^2}\}= \{1, \frac{\sin^2(R)}{R^2}, \frac{\sin^2(R)}{R^2}\}$, where we've used $\sin^2(-x) = \sin^2(x)$. The determinant of $g$ is just the product of its eigenvalues, so plugging this into \eqref{dV}, we find $dV = \frac{\sin^2(R)}{R^2} dx dy dz$. Let us formally change to Polar coordinates, $x=R\sin(\theta) \cos(\phi), y=R\sin(\theta) \sin(\phi)$, and $z= R\cos(\theta)$. This comes with the usual Jacobian of $R^2 \sin(\theta)$. The volume element is

\begin{align}
   dV =\frac{\sin(R)^2}{R^2} R^2 dR \sin(\theta) d\theta d\phi = \sin(R)^2 dR \sin(\theta) d\theta d\phi. \label{SU2VolumeAlgebra}
\end{align}

The final step is to define linear variables $f(R)$ and $t(\theta)$, which we can sample uniformly in the correct range, and then solve for $(R,\theta)$ to get the correct weight. The $\theta$ variable is straightforward: Define $\theta= \arccos(t)$, and sample $t$ uniformly in $[-1,1].$ We can now sample $f \in [0,f(\pi)]$ uniformly, and solve for $R$ numerically.

\begin{align}
    f(R)&= \int_0^R \sin(R')^2 dR' = \frac{1}{2} \int_0^R (1-\cos(2R')) dR' = (R - \sin(R)\cos(R))/2
\end{align}

To summarize: To get the Haar measure on $SU(2)$ in terms of the parameters $(x,y,z)$ in $U=e^{i(x X+yY+zZ)}$, we sample $(f,t,\phi)$ uniformly in $[0,f(\pi)]$, $[-1,1]$, and $[0,2\pi]$ respectively, solve $f(R)=(R - \sin(R)\cos(R))/2$ numerically to find $R$, and define $\theta= \arccos(t)$. This gives us the correct distribution $(R,\theta,\phi)$ in polar coordinates. Finally, we convert back to Cartesian coordinates to get the coefficients $(x,y,z)$.

\end{document}